\documentclass[11pt, oneside]{article}   	% use "amsart" instead of "article" for AMSLaTeX format
\usepackage[margin=1.1in]{geometry}                		% See geometry.pdf to learn the layout options. There are lots.
\pdfoutput=1
\usepackage{graphicx}				% Use pdf, png, jpg, or eps§ with pdflatex; use eps in DVI mode
\usepackage{amssymb}
\usepackage{biblatex}
\usepackage{makecell}
\usepackage{hyperref}
\usepackage{multirow}
\usepackage{xcolor}
\usepackage{enumitem}
\setlist[itemize]{noitemsep}
\newcommand{\E}{\mathbb{E}}
\newcommand{\Var}{\mathrm{Var}}
\newcommand{\Cov}{\mathrm{Cov}}
\newcommand{\Corr}{\mathrm{Corr}}
\newcommand{\standarderror}{\mathrm{SE}}
\newcommand{\confidenceinterval}{\mathrm{CI}}

\hypersetup{
  colorlinks=true,
  linkcolor=blue,
  filecolor=magenta,
  urlcolor=black,
  pdftitle={Adding Error Bars to Evals: A Statistical Approach to Language Model Evaluations},
  pdfauthor={Evan Miller},
  pdfpagemode=FullScreen,
  citecolor=teal,
}
\title{Adding Error Bars to Evals: \\ A Statistical Approach to Language Model Evaluations}
\author{Evan Miller \\ Anthropic \\ \href{mailto:evanmiller@anthropic.com}{\texttt{evanmiller@anthropic.com}}}
\begin{document}

\maketitle

\begin{abstract}
Evaluations are critical for understanding the capabilities of large language models (LLMs). Fundamentally, evaluations are experiments; but the literature on evaluations has largely ignored the literature from other sciences on experiment analysis and planning. This article shows researchers with some training in statistics how to think about and analyze data from language model evaluations. Conceptualizing evaluation questions as having been drawn from an unseen super-population, we present formulas for analyzing evaluation data, measuring differences between two models, and planning an evaluation experiment. We make a number of specific recommendations for running language model evaluations and reporting experiment results in a way that minimizes statistical noise and maximizes informativeness.
\end{abstract}

\section{Introduction}
\label{introduction}

Language models are measured in the literature by evaluations, or evals. Evals are commonly run and reported with a ``highest number is best'' mentality; industry practice is to highlight a state-of-the-art (SOTA) result in bold, but not necessarily to test that result for any kind of statistical significance.\cite{madaan2024quantifyingvarianceevaluationbenchmarks} Chatbot Arena\cite{chiang2024chatbot} has popularized the use of confidence intervals in its Elo scores, but error bars remain noticeably absent from traditional question-and-answer evals. One recent and notable exception is the technical report on the Llama 3 model family\cite{dubey2024llama3herdmodels}, which includes simple confidence intervals on a number of evals.

In this article, we seek to introduce rigorous statistical thinking into the world of language model evaluations, so that researchers may quantify the precision with which they are able to answer questions and test hypotheses using evals. After developing a comprehensive analytic framework, we make specific recommendations for the computation of confidence intervals and the reporting of eval results. Using this framework, we show that the confidence intervals recently reported in \cite{dubey2024llama3herdmodels} are likely too narrow in some cases and too wide in other cases.

A short hypothetical example will motivate the overall discussion. Imagine that two competing models, code-named ``Galleon'' and ``Dreadnought'', are being considered for deployment in a particular application (say, with a bent toward coding and mathematical reasoning tasks). As part of the decision-making process, three popular language model evaluations are performed on the two models: MATH, a mathematical reasoning eval\cite{hendrycks2021math}; HumanEval, a Python coding eval\cite{chen2021evaluatinglargelanguagemodels}; and MGSM, a multilingual eval covering grade-school math\cite{shi2022languagemodelsmultilingualchainofthought}. The fictional results from this hypothetical comparison are presented in Table \ref{table:1}.

\begin{table}
\centering
\begin{tabular}{ |c|>{\centering\arraybackslash}p{0.16\linewidth}|>{\centering\arraybackslash}p{0.16\linewidth}|c| }
\hline
Eval \textbackslash \ Model & ``Galleon'' & ``Dreadnought'' & Difference \\
\hline
\makecell[c]{MATH} & 65.5\% & 63.0\% & $+2.5\%$\\
\hline
\makecell[c]{HumanEval} & 83.6\% & 87.7\% & $-3.1\%$\\
\hline
\makecell[c]{MGSM} & 75.3\% & 78.0\% & $-2.7\%$\\
\hline
\end{tabular}
\caption{Hypothetical data from two fictional models across three (non-fictional) evals}
\label{table:1}
\end{table}

On its face, the data table presents conflicting results: Galleon appears to outperform Dreadnought on MATH (65.5\% vs 63.0\%), but Dreadnought has performed better on HumanEval and MGSM by slightly wider margins. Is it safe to conclude from the results that Dreadnought is generally better suited for coding and mathematical tasks, given its margin of victory on two of three evals? Or should something in the data potentially give us pause?

``Evaluating the evaluations'' is a complex undertaking fraught with both qualitative and quantitative considerations.\cite{ganguli2023challenges} Remaining agnostic about the relative and qualitative merits of various evals, this article develops a framework for answering quantitative questions about specific eval results. With the aim of informing holistic decision-making, we offer a series of recommendations for running and reporting evals in a way that enables researchers to test well-formed hypotheses about competing models, competing hyperparameters, and competing prompts. Our specific recommendations to researchers include:

\begin{enumerate}
\item Computing standard errors of the mean using the Central Limit Theorem
\item When questions are drawn in related groups, computing clustered standard errors
\item Reducing variance by resampling answers and by analyzing next-token probabilities
\item When two models are being compared, conducting statistical inference on the question-level paired differences, rather than the population-level summary statistics
\item Using power analysis to determine whether an eval (or a random subsample) is capable of testing a hypothesis of interest
\end{enumerate}

Drawing on statistical theory and the experimental design literature, we demonstrate that a small number of conceptual assumptions unlocks a rich theoretical landscape for researchers studying language model evaluations, and show practitioners how to conduct statistical inference on often-noisy eval data. For an overview of experiment design, we refer the reader to \cite{Imbens_Rubin_2015}.

\section{Analysis framework}

Suppose that the questions in an eval do not represent all possible questions, but instead were drawn at random from a (hypothetical, infinite, unseen) super-population of questions. This simple supposition lets us jump ``through the looking glass'' of the specific questions that appear in an eval in order to study the underlying {\em skill} that the eval is attempting to measure. We modify this assumption in Section \ref{sec:clustered_qs} to study questions that may have been drawn together.

\subsection{Independent questions}

More formally, suppose an eval consists of $n$ independently drawn questions. We write the score on question $i$ as $s_i$, decomposing the score into a mean component $x_i$ and a zero-mean random component $\epsilon_i$:
\[
s_i = x_i + \epsilon_i
\]

We refer to $x_i$ as the {\em conditional mean} and the variance of the random component $\epsilon_i$ as the {\em conditional variance}, that is, the mean and variance conditional on the selection of question $i$ in the eval. Denote this latter quantity $\sigma_i^2=\Var(\epsilon_i)$.

We also write an unconditional version of these numbers (that is, unconditional on the selection of $i$) as
\[
s = x + \epsilon
\]

Let $\mu$ represent the unobserved mean value of the super-population with $\mu=\E[s]=\E[x]$. We wish to conduct inference on  (that is, the ``true'' mean eval score) given only the observed question scores $s_1,\ldots,s_n$. Let $\bar{s}=\frac{1}{n}\sum_i s_i$ represent the average of the observed scores. It follows from the Law of Large Numbers that $\mu$ may be estimated using $\hat{\mu}=\bar{s}$, and from the Central Limit Theorem (C.L.T.) that the standard error of $\hat{\mu}$ can be estimated as
\begin{equation}
\label{eqn:se_clt}
\standarderror_{\rm C.L.T.} = \sqrt{\Var(s)/n}=\sqrt{\left(\frac{1}{n-1}\sum_i(s_i-\bar{s})^2\right)/n}
\end{equation}

In the special case where $s_i \in \{0, 1\}$ (a Bernoulli variable), Equation \ref{eqn:se_clt} becomes

\begin{equation}
\label{eqn:se_bernoulli}
\standarderror_{\rm Bernoulli} = \sqrt{\bar{s}(1-\bar{s})/n}
\end{equation}

We note that it is common practice to compute the standard error of the mean by bootstrapping; see, for instance, the OpenAI evals\cite{openai-evals} frameworks. But the Central Limit Theorem is applicable to any evals having scores with finite variance and a large number of questions, and so we regard bootstrapping as unnecessary unless a complicated sampling scheme or estimator is being used. We also note that \cite{dubey2024llama3herdmodels} incorrectly estimates all of its standard errors using $\standarderror_{\rm Bernoulli}$, even when $s_i$ takes fractional values, such as with an F1 score. In these cases, $\standarderror_{\rm Bernoulli}$ will tend to be conservative (too wide) compared to $\standarderror_{\rm C.L.T.}$. The Inspect framework\cite{uk-inspect} correctly computes $\standarderror_{\rm C.L.T.}$ with its built-in \texttt{stderr()} metric.

We suggest reporting the standard error of the mean alongside (beneath) the mean when reporting eval scores. A common practice in other sciences is to report the standard error in parentheses; we suggest emulating this practice. See Table \ref{faketable:clt} for an example.

\begin{table}
\centering
\begin{tabular}{ |c|c|>{\centering\arraybackslash}p{0.2\linewidth}|>{\centering\arraybackslash}p{0.2\linewidth}| }
\hline
\  & \# Questions & ``Galleon'' & ``Dreadnought'' \\
\hline
\makecell[c]{MATH} & 5,000 & \makecell{65.5\% \\ (0.7\%)} & \makecell{63.0\% \\ (0.7\%)} \\
\hline
\makecell[c]{HumanEval} & 164 & \makecell{83.6\% \\ (3.2\%)} & \makecell{86.7\% \\ (3.0\%)} \\
\hline
\makecell[c]{MGSM} & 2,500 & \makecell{75.3\% \\ (0.9\%)} & \makecell{78.0\% \\ (0.9\%)} \\
\hline
\end{tabular}
\caption{We suggest two new reporting practices: including the number of questions in each eval, and the standard error of each estimate in parentheses (fictional models and numbers).}
\label{faketable:clt}
\end{table}

A 95\% confidence interval may be computed from such a table as
\begin{equation}
\label{eqn:ci_clt}
\confidenceinterval_{\ 95\%} = \bar{s} \pm 1.96 \times \standarderror_{\rm C.L.T.}
\end{equation}

\subsection{Clustered questions}
\label{sec:clustered_qs}

We next consider eval questions that are drawn in groups, or clusters. For instance, DROP\cite{dua-etal-2019-drop}, QuAC\cite{choi2018quacquestionanswering}, RACE\cite{lai-etal-2017-race}, and SQuAD\cite{rajpurkar2018knowdontknowunanswerable} are reading-comprehension evals having multiple related questions about independently selected passages of text, and multilingual evals such as MGSM\cite{shi2022languagemodelsmultilingualchainofthought} consist of the same question translated into many languages. Because the inclusion of questions is non-independent, a key assumption of the Central Limit Theorem (or a bootstrap) is violated, and so a naive application of Equation \ref{eqn:se_clt} will yield inconsistent standard errors. Here we show how to use clustered standard errors\cite{clustered}, a technique developed in the social sciences, to account for the dependence and correlation structure present in question clusters.

Let $s_{i,c}$ denote the $i$th question score within cluster $c$, and assume that draws of clusters are independent. Continue to denote the mean observed score as $\bar{s}$. The cluster-adjusted standard error of the mean score can be computed as

\begin{equation}
\label{eq:clustered}
\standarderror_{\rm clustered} = \left(\standarderror_{\rm C.L.T.}^2 + \frac{1}{n^2}\sum_c \sum_i \sum_{j\ne i} (s_{i, c}-\bar{s})(s_{j, c}-\bar{s})\right)^{1/2}
\end{equation}

The clustered standard error acts as a kind of ``sliding scale'' between cases where scores within a cluster are perfectly correlated (in which case each cluster acts as a single independent observation) and perfectly uncorrelated (in which case the clustered standard error is equivalent to the unclustered case). The intra-cluster correlations (or lack thereof) are captured by the triple summation (over clusters and cross-terms within clusters); for a derivation of Equation \ref{eq:clustered}, see Appendix \ref{apx:clustered}.

\begin{table}
\centering
\begin{tabular}{ |c|c|c|>{\centering\arraybackslash}p{0.16\linewidth}|>{\centering\arraybackslash}p{0.16\linewidth}| }
\hline
\  & \# Questions & \# Clusters & ``Galleon'' & ``Dreadnought'' \\
\hline
\makecell[c]
{DROP} & 9,622 & 588 & \makecell{87.1 \\ (0.8)} & \makecell{83.1 \\ (0.9)} \\
\hline
RACE-H & 3,498 & 1,045 & \makecell{91.5\% \\ (0.5\%)} & \makecell{82.9\% \\ (0.7\%)} \\
\hline
{MGSM} & 2,500 & 250 & \makecell{75.3\% \\ (1.6\%)} & \makecell{78.0\% \\ (1.5\%)} \\
\hline
\end{tabular}
\caption{We suggest including the cluster count alongside the question count when reporting cluster-adjusted standard errors (fictional models and numbers).}
\label{table:clustered_fake}
\end{table}

\begin{table}
\centering
\begin{tabular}{ |c|>{\centering\arraybackslash}p{0.16\linewidth}|>{\centering\arraybackslash}p{0.16\linewidth}|>{\centering\arraybackslash}p{0.16\linewidth}| }
\hline
\  & $\standarderror_{\rm clustered}$ & $\standarderror_{\rm C.L.T.}$  & Ratio \\
\hline
\makecell[c]
{DROP} & (1.34) & (0.44) & 3.05 \\
\hline
RACE-H & (0.51\%) & (0.46\%) & 1.10 \\
\hline
{MGSM} & (1.62\%) & (0.86\%)  & 1.88 \\
\hline
\end{tabular}
\caption{Clustered and naive standard errors computed on two popular evals using Anthropic models (non-fictional numbers). Analyzing the same data, clustered standard errors can be over 3X larger than naive standard errors.}
\label{table:clustered_real}
\end{table}

A suggested format for reporting cluster-adjusted standard errors is presented in Table \ref{table:clustered_fake}, and some real-world numbers are reported in Table \ref{table:clustered_real}. We note that the cluster adjustment in our real-world example is far from trivial (up to 3X). Failure to adjust standard errors for clustered sampling may lead an unsuspecting analyst to suppose that the measurement of the overall eval score is much more precise than it actually is. We therefore advise that the confidence intervals for reading-comprehension evals reported in \cite{dubey2024llama3herdmodels} are likely anti-conservative (too narrow).

\section{Variance reduction}

The standard error of $\hat{\mu}$ quantifies the uncertainty associated with our estimate of the overall eval score. Reducing this quantity (which is the square root of the estimate variance) improves the precision of the estimate.

The variance associated with $\hat{\mu}$ may be decomposed into two components: the {\em variance of the conditional mean}, that is, the variance associated with choosing questions from the super-population, and the {\em mean conditional variance}, which is the mean variance of scores associated with the questions that were chosen. This decomposition is additive, and follows from the law of total variance. Mathematically,

\[
\Var(\hat{\mu})=\Var(s)/n=(\Var(x)+\E[\sigma_i^2])/n
\]

This equation has a few implications: the simplest way to reduce the variance of $\hat{\mu}$ is to increase $n$, the number of sampled questions. The variance of the conditional mean, $\Var(x)$, is a property of the super-population and therefore immutable; but we have a couple of strategies available for reducing the overall variance via the expected conditional variance, $\E[\sigma_i^2]$.

\subsection{Resampling}
\label{sec:resampling}

The first strategy for reducing the expected conditional variance is to resample the model a number of times, and to compute the standard error using the question-level mean scores from the resamples. Suppose each question is sampled (answered) $K$ times, and the score $s_i$ is the mean of these $K$ answer scores. Since the conditional variances are equal for all $K$ answer scores, the overall conditional variance becomes

\[
\Var(s_i)=\sigma_i^2/K
\]

This relation should clarify the issue of how many times to resample a given question. Once $\E[\sigma_i^2]/K \ll \Var(x)$, increasing $K$ further will have little effect on the standard error of $\hat{\mu}$.

We work through an example to show how to compute a value for $K$. Suppose scores are binary (0 or 1) and question difficulty is uniformly distributed, $x \sim U[0, 1]$. Then $\epsilon_i = 1-x_i$ with probability $x_i$ and $\epsilon_i=-x_i$ otherwise. A bit of integration reveals that $\Var(x)=1/12$ and $\E[\sigma_i^2]= 1/6$. The required relation reduces to $K\gg2$, or equivalently $2/K\ll1$. Writing the variance of this estimator with arbitrary $K$ in terms of the estimator with $K=1$, we have

\[
\Var(\hat{\mu}|K>1)=\Var(\hat{\mu}|K=1)\times (1+2/K)/3
\]

Going from $K=1$ (no resampling of answers) to $K=2$, the total variance is reduced by 1/3. Increasing to $K=4$, we have a variance reduction of 1/2, and setting $K=6$, we reduce variance by 5/9. The upper limit on variance reduction via resampling in this example is 2/3.

Note that computing a pooled standard error across all $KN$ answers will be inconsistent, as multiple answers to the same question would violate the assumption of independent draws. Refer to Section \ref{sec:clustered_qs} for a discussion of questions drawn in related groups.

\subsection{Next-token probabilities}

The second strategy for reducing the expected conditional variance $\E[\sigma_i^2]$ is to eliminate the term altogether. For language model evals that do not utilize chain-of-thought reasoning, the conditional variance can be removed by analyzing next-token probabilities, rather than evaluating the model’s sampled (or resampled) output.

Consider for instance a multiple-choice eval, and a prompt that induces a model to produce its answer in its first token. If a correct answer is worth 1 and an incorrect answer is worth 0, and the probability of the correct token is denoted $p_i$, then $s_i=x_i=p_i$ and $\epsilon_i=0$. This yields $\Var(\hat{\mu})=\Var(p)/n$. Using the uniform-difficulty example from the previous section, next-token probabilities will reduce the variance of the estimator by 2/3 (the upper limit achievable via resampling) compared to grading a single sample from each question.

\subsection{Don't touch the thermostat!}

It may be tempting to reduce the ``sampling temperature''\cite{hinton2015distillingknowledgeneuralnetwork} of the model in order to reduce (or eliminate) the conditional variance. However, we advise against this practice, unless the purpose is to study the model at the new temperature. Besides altering the model’s behavior, adjusting the sampling temperature may simply shift the conditional variance (which can be mitigated using the two techniques above) into the variance of the conditional means (which cannot), or else reduce conditional variance by injecting bias into the estimator. Two short examples will illustrate these points.

Consider a single-token true/false eval where the conditional score means at $T=1$ are uniformly distributed, $x_{T=1}\sim U[0, 1]$. As in Section \ref{sec:resampling}, $\Var(x_{T=1})=1/12$. But at $T=0$, $x_{T=0}=1\{x_{T=1}>0.5\}$ and the uniform distribution is ``rounded'' into a Bernoulli distribution with $p=1/2$. So $\Var(x_{T=0})=1/4$. In this case, reducing the sampling temperature, and thereby eliminating the conditional variance, has inadvertently tripled the minimum variance in the score data from 1/12 to 1/4.

In the above case, $\E[x_{T=1}]=\E[x_{T=0}]$, but this does not always hold. Consider a similar (single-token, true/false) case where $x_{T=1}\sim U[1/3,1]$ and (as a consequence) $x_{T=0}$ is rounded to a Bernoulli distribution with $p=1/4$. Then $\E[x_{T=1}]=2/3<\E[x_{T=0}]=3/4$ and  $\Var(x_{T=1})=1/27 \ll \Var(x_{T=0})=3/16$; that is, not only has the temperature change shifted the expected score, but the variance of the conditional means has increased approximately five-fold.

We therefore recommend a two-pronged variance-reduction strategy. When next-token probabilities are available, and the language model eval can be conducted using next-token probabilities (i.e. without token generation), compute the expected score for each question, and compute the standard error of expected scores across questions. When next-token probabilities are not available, or the answer requires a chain of thought or other complex interaction, choose a $K$ such that $\E[\sigma_i^2]/K \ll \Var(x)$ and compute the standard error across question-level mean scores. In neither case should the sampling temperature be adjusted for the sake of reducing variance in the scores.

\section{Comparing models}

Thus far we have only analyzed the standard error of eval scores considered in isolation. But a particular model’s score on a given eval usually does not have any inherent meaning; it primarily makes sense in relation to the scores of other models. In this section we provide formulas for comparing the scores of two models so that an analyst might determine if a model is outperforming another model in a statistically significant way, or if the difference between two models is indistinguishable from noise.

\subsection{Unpaired analysis}

We introduce model subscripts $A$ and $B$ for the remainder of the article. A naive comparison between eval scores can be made by computing the difference between mean eval scores
\[
\hat{\mu}_{A-B}=\hat{\mu}_A-\hat{\mu}_B
\]

and an associated standard error
\[
\standarderror_{A-B}=\sqrt{\standarderror_A^2+\standarderror_B^2}
\]

This two quantities can be used to compute the usual 95\% confidence interval and z-score
\begin{equation}
\label{eqn:unpaired_ci}
\confidenceinterval_{A-B, 95\%}=\hat{\mu}_{A-B} \pm 1.96 \times \standarderror_{A-B}
\end{equation}
\begin{equation}
\label{eqn:unpaired_z}
z_{A-B}=\hat{\mu}_{A-B}/\standarderror_{A-B}
\end{equation}

If two models independently report their eval scores and standard errors, it is possible for an analyst to test their difference for statistical significance -- even if the two model reports used non-identical random subsets of eval questions.

\subsection{Paired analysis}
\label{sec:paired}

The naive comparison above misses an opportunity to reduce the standard error when two models evaluate the same set of questions. Let $s_{A-B, i}=s_{A, i}-s_{B, i}$ represent the difference between scores on question $i$, and let $\bar{s}_{A-B}=\bar{s}_A-\bar{s}_B$ represent the observed average difference. Then we can estimate the standard error of the estimated difference as
\begin{equation}
\standarderror_{A-B, {\rm paired}} = \sqrt{\Var(s_{A-B})/n} = \sqrt{\left(\frac{1}{n-1}\sum_i (s_{A-B, i}-\bar{s}_{A-B})^2\right)/n}
\end{equation}

This revised standard error can be plugged into Equations \ref{eqn:unpaired_ci} and \ref{eqn:unpaired_z} to compute a confidence interval and z-score.

We can compute the reduction in variance achieved with this paired differences test over the unpaired test. First, write out an expression for the variance in the unpaired case
\[
\Var(\hat{\mu}_{A-B, {\rm unpaired}})=(\Var(s_A)+\Var(s_B))/n
\]
and the paired case
\[
\Var(\hat{\mu}_{A-B, {\rm paired}})=(\Var(s_A)+\Var(s_B)-2\ \Cov(s_A, s_B))/n
\]
Combining, and recognizing that the cross-model residuals are uncorrelated, we have
\[
\Var(\hat{\mu}_{A-B, {\rm paired}})=\Var(\hat{\mu}_{A-B, {\rm unpaired}})-2\ \Cov(x_A, x_B)/n
\]

We can thus reduce the variance with paired differences as long as the conditional means of the model scores are correlated; that is to say, if the two models have some amount of agreement on which questions are ``easy'' and which questions are ``hard''.

A short calculation will demonstrate the degree of variance reduction that might be expected. Suppose an eval uses next-token probabilities to form continuous scores with zero conditional variance, and that these scores are uniformly distributed for two models over the $[0, 1]$ interval. Suppose that the model scores have a correlation coefficient of 0.5. Then $\Var(s_A)=\Var(s_B)=1/12$ and $\Cov(x_A, x_B)=0.5\sqrt{\Var(s_A)\Var(s_B)}=1/24$. In this case, using paired differences will reduce the variance of the estimator by 1/3 in relative terms (that is, from 1/6 to 1/9 in absolute terms).

Because eval question scores are likely to be positively correlated, even across unrelated models, paired differences represent a ``free'' reduction in estimator variance when comparing two models. We therefore recommend using the paired version of the standard error estimate wherever practicable. We encourage authors of technical reports to include pairwise differences, pairwise standard errors, and score correlations whenever two or more models are being evaluated. Pairwise standard errors may be computed either directly on the differences, or using the single-sample standard errors, the Pearson product-moment correlation, and the relation
\[
\standarderror_{A-B, {\rm paired}}=\sqrt{\standarderror_A^2+\standarderror_B^2-2\ \standarderror_A \standarderror_B \Corr(s_A, s_B)}
\]

A clustered version of the standard error, appropriate for DROP, QuAC, RACE, SQuAD, MGSM, and other evals where questions are drawn in related groups, is directly computable from the differences as
\begin{equation}
\standarderror_{A-B, {\rm paired}, {\rm clustered}}=\frac{1}{n}\left(\sum_c \sum_i \sum_j (s_{A-B, i, c}-\bar{s}_{A-B})(s_{A-B, j, c}-\bar{s}_{A-B})\right)^{1/2}
\end{equation}

where $s_{A-B, i, c}$ denotes the score difference on the $i$th question within cluster $c$.

A suggested table format for reporting pairwise results is provided in Table \ref{faketable:pairwise}. A 95\% confidence interval on model differences may be computed from the base estimate and standard error, as in Equation \ref{eqn:ci_clt}.

\begin{table}
\begin{center}
\begin{tabular}{ |c|c|c|c|c|c| }
\hline
Eval & Model & Baseline & Model -- Baseline & 95\% Conf. Interval & Correlation \\
\hline
MATH & Galleon & Dreadnought & +2.5\% (0.7\%) & (+1.2\%, +3.8\%) & 0.50 \\
\hline
HumanEval & Galleon & Dreadnought & $-3.1\%$ (2.1\%) & $(-7.2\%, +1.0\%)$ & 0.64 \\
\hline
MGSM & Galleon & Dreadnought & $-2.7\%$ (1.7\%) & $(-6.1\%, +0.7\%)$ & 0.37 \\
\hline
\end{tabular}
\caption{Suggested presentation of pairwise differences, standard errors, confidence intervals, and correlation values as a supplement to main results. In the fictional data above, the difference between the two models on MATH is statistically significant (the confidence interval is positive), but the differences on HumanEval and MGSM are not statistically significant at the 5\% level.}
\label{faketable:pairwise}
\end{center}
\end{table}

We now possess the analytic tools needed to rigorously answer the questions posed in the Introduction. Using pairwise analysis on all three evals, and ensuring that standard errors were appropriately clustered on MSGM, the numbers in Table \ref{faketable:pairwise} would lead us to conclude that the Galleon indeed outperformed Dreadnought on MATH in a statistically significant way – but that the differences on HumanEval and MGSM are indistinguishable from statistical noise. In other words, while a superficial reading of the eval data might have originally tempted us to conclude that Dreadnought was the overall better-performing model, a closer examination of the data would tend to lead the careful analyst to the opposite conclusion.

\section{Power analysis}

{\em Power} refers to the ability of an experiment to make a measurement of interest in the presence of statistical noise.\cite{NBERw15701} In the context of language model evals, we may wish to know whether a model represents an improvement of some magnitude over another model.\cite{card2020littlepowercomesgreat} Due to the variance implied by sampling questions from the super-population (plus the conditional variance after the questions are chosen), power must always be defined in terms of probability. In this section we present a sample-size formula needed to conduct power analysis for language model evals, and apply it in a worked example to answer the empirical question posed in Section \ref{introduction}.

The sample-size formula -- describing the relationship between the hypothesized difference between two models and the number of questions included in an experiment -- ought to prove useful in several ways. Consumers of existing evals may use the formula to determine the number of questions to subsample from a large eval, or to determine an appropriate value of $K$ defined in Section \ref{sec:resampling}. If the number of questions in the eval is fixed, consumers can calculate the Minimum Detectable Effect and decide whether the eval is worth running. The authors of new evals may use the formula to decide how many questions should be commissioned.

The inputs into the sample-size formula include:

\begin{itemize}
\item Significance level $\alpha$, which represents the Type I error rate under the null hypothesis
\item Power level $1-\beta$, where $\beta$ represents the Type II error rate under the alternative hypothesis
\item Minimum Detectable Effect $\delta$, which represents the mean score difference between two models under the alternative hypothesis
\end{itemize}

To simplify notation, let
\[
\omega^2 = \Var(x_A) + Var(x_B) - 2\Cov(x_A, x_B)
\]
\[
\sigma_A^2=\E[\sigma_{A, i}^2]
\]
\[
\sigma_B^2=\E[\sigma_{B, i}^2]
\]

Let $z_p$ represent the $(1-p)$th percentile of a standard normal distribution. We assume a paired analysis described in Section \ref{sec:paired}, and that answers will be sampled $K_A$ times from model $A$ and $K_B$ times from model $B$ (in the simplest case, $K_A=K_B=1$). Then the number of independent questions $n$ required to achieve a Type I error rate $\alpha$ and Type II error rate $\beta$ with a given Minimum Detectable Effect $\delta$ is
\begin{equation}
\label{eqn:sample_size}
n=(z_{\alpha/2}+z_\beta)^2(\omega^2+\sigma_A^2/K_A + \sigma_B^2/K_B)/\delta^2
\end{equation}

The quantities $\omega^2$, $\sigma_A^2$, and $\sigma_B^2$ may be estimated from previous eval data. A short derivation of the above formula is presented in Appendix \ref{apx:sample_size}.

As a simple example, suppose $\sigma_A^2=\sigma_B^2=0$ and $\omega^2=1/9$, following the conditions described in Section \ref{sec:paired}. Suppose we wish to detect an absolute difference of $\delta=0.03$ at least 80\% of the time ($\beta=0.20$) with a false-positive rate of 5\% ($\alpha=0.05$). Then the eval will need to contain at least
\[
n=(z_{0.025}+z_{0.20})^2(1/9)/(0.03)^2 \approx 969
\]

independent questions. Although these parameters are fictional, they are reasonable, and suggest that new evals should contain at least 1,000 questions in order to have good signaling ability.

If the number of questions is fixed, and the practitioner wishes to know the Minimum Detectable Effect associated with $n$, Equation \ref{eqn:sample_size} is easily inverted as
\begin{equation}
\label{eqn:mde}
\delta = (z_{\alpha/2}+z_B)\sqrt{(\omega^2+\sigma_A^2/K_A+\sigma_B^2/K_B)/n}
\end{equation}

The above equation may be used, for instance, to predict the effect of increasing the per-question sample counts $K_A$ and $K_B$ on the Minimum Detectable Effect in a nondeterministic eval. Suppose that $\sigma_A^2=\sigma_B^2=1/6$ and $\omega^2=1/9$, following the conditions of Section \ref{sec:resampling} with an additional assumption that $\Corr(x_A, x_B)=0.5$. Suppose $n=198$, $\alpha=0.05$, and $\beta=0.20$. It follows from Equation \ref{eqn:mde} that increasing $K_A=K_B$ from 1 to 10 reduces the Minimum Detectable Effect from 13.2\% to 7.5\%.

Cluster-adjusted versions of Equations \ref{eqn:sample_size} and \ref{eqn:mde} are included in Appendix \ref{apx:cluster_sample}.

\section{Conclusion}

This article has presented a statistical treatment of language model evaluations, drawing heavily from existing literature in experiment design.

For single-model analysis, we presented analytic formulas for naive and clustered standard errors, and for two-model analysis, we presented formulas for unpaired, paired, and paired-and-clustered standard errors. We recommended several techniques for reducing the variance of estimates, including resampling answers, analyzing next-token probabilities, and computing question-level differences between models, and advised against adjusting the sampling temperature for the sake of variance reduction. We suggest that practitioners include standard errors of their eval scores in their technical reports, and also include pairwise differences, pairwise standard errors, and score correlations when multiple models are being compared. We presented a sample-size formula so that model evaluators can determine in advance the size of difference that may be reliably detected between two models on a given eval using a given resampling strategy.

Experiment design represents a large and venerable literature. We hope that with proper statistical tools, such as those presented in this article, machine learning practitioners will think of their model evaluations as informative experiments rather than a series of contests to produce the largest number. We encourage researchers to continue exploring statistical techniques found in other experimental fields in order to further enrich our shared understanding of language models and their capabilities.

\printbibliography

\newpage

\appendix

\section{Clustered standard errors}
\label{apx:clustered}

We approach the problem with linear regression. Let $s_{i,c}$ denote the $i$th of $n_c$ question scores within cluster $c$, decomposed into a mean and random component as

\[
s_{i,c} = x_{i,c} + \epsilon_{i,c}
\]

Let $\delta_{i,c}=x_{i,c}-\mu$ represent the deviation of the conditional (question-level) mean from the true mean (that is, the hypothetical mean across all questions and clusters). Then the regression can be specified as

\[
s_{i,c}=\mu + \delta_{i,c} + \epsilon_{i,c}
\]

where $\epsilon_{i,c}$ is a random component and $\delta_{i,c}$ acts as a question-level fixed effect that is not separately estimated. We continue to estimate $\hat{\mu}=\bar{s}$ and denote the regression residual $u_{i,c} = s_{i,c}-\bar{s}$. The traditional clustered standard error formula is

\[
\Var_{\rm clustered}(\hat{\mu})=(X'X)^{-1}\left(\sum_c X_c' \Omega X_c\right)(X'X)^{-1}
\]

where $X$ represents the full matrix of covariates, $X_c$ represents the covariates within cluster $c$, and $\Omega_c$ represents the residual covariance matrix within cluster $c$.

In our application, $X=1_n$ (a vector of $n$ 1s), $X_c=1_{n_c}$ (a vector of $n_c$ 1s), and $\Omega_c = u_c u_c'$. Thus $(X'X)^{-1}=1/n$ and $X_c'\Omega_c X_c=\sum_i \sum_j u_{i,c} u_{j,c}=\sum_i \sum_j(s_{i,c}-\bar{s})(s_{j,c}-\bar{s})$. Plugging in these values yields

\[
\Var_{\rm clustered}(\hat{\mu})=\sum_c \sum_i \sum_j (s_{i,c}-\bar{s})(s_{j,c}-\bar{s})/n^2=\sum_c \sum_i(s_{i,c}-\bar{s})^2/n^2+\sum_c \sum_i \sum_{j\ne i} (s_{i,c}-\bar{s})(s_{j,c}-\bar{s})/n^2
\]

Recognizing that the first term is equal to the unclustered variance, we can write

\[
\Var_{\rm clustered}(\hat{\mu})=\Var_{\rm unclustered}(\hat{\mu})+\sum_c \sum_i \sum_{j\ne i} (s_{i,c}-\bar{s})(s_{j,c}-\bar{s})/n^2
\]

The two-sample version can be developed by analyzing the question-level score differences rather than the scores.

\section{Sample-size formula derivation}
\label{apx:sample_size}

Following \cite{NBERw15701}, we set up the power analysis with a hypothetical measurement $\tilde{s}_{A-B}$ that will trigger a Type I error with probability $\alpha$ and a Type II error rate with probability $\beta$. The z-scores on such a measurement under the null hypothesis and the alternative hypothesis are

\[
z_{\alpha/2}=\tilde{s}_{A-B}/\sqrt{(\omega^2+\sigma_A^2/K_A+\sigma_B^2/K_B)/n}
\]
\[
z_\beta = (\delta - \tilde{s}_{A-B})/\sqrt{(\omega^2 + \sigma_A^2/K_A + \sigma_B^2/K_B)/n}
\]

Combining to eliminate $\tilde{s}_{A-B}$, we have an expression for the MDE in terms of the other variables

\[
\delta = (z_{\alpha/2}+z_B)\sqrt{(\omega^2+\sigma_A^2/K_A+\sigma_B^2/K_B)/n}
\]

Or inverting the equation, we have a sample-size formula for the number of questions $n$ required to produce a desired MDE $\delta$

\[
n=(z_{\alpha/2}+z_\beta)^2(\omega^2+\sigma_A^2/K_A + \sigma_B^2/K_B)/\delta^2
\]

\section{Cluster-adjusted sample-size formula}
\label{apx:cluster_sample}

In order to account for clustered questions, the sample-size formula requires cluster-adjusted versions of $\omega^2$, $\sigma_A^2$, and $\sigma_B^2$. In this section we develop formulas for estimating these quantities from previous eval data.

Start with the clustered score variance estimator

\[
\Var_{\rm clustered}(\hat{\mu}_{A-B})=\frac{1}{n}\sum_c \sum_i \sum_j (s_{A-B, i, c} - \bar{s}_{A-B})(s_{A-B, j, c}-\bar{s}_{A-B})
\]

we can decompose $s$ into $x$ and $\epsilon$,

\[
\Var_{\rm clustered}(\hat{\mu}_{A-B}) = \frac{1}{n}\sum_c \sum_i \sum_j (x_{A, i, c} - x_{B, i, c} + \epsilon_{A, i, c} - \epsilon_{B, i, c} - \bar{s}_{A-B})(x_{A, j, c} - x_{B, j, c} + \epsilon_{A, j, c} - \epsilon_{B, j, c}-\bar{s}_{A-B})
\]

which, dropping cross-terms which are zero in expectation, will reduce to

\[
\Var_{\rm clustered}(\hat{\mu}_{A-B}) = \frac{1}{n}\sum_c \sum_i \sum_j (x_{A, i, c} - x_{B, i, c} - \bar{s}_{A-B})(x_{A, j, c} - x_{B, j, c} -\bar{s}_{A-B})
\]
\[ + \frac{1}{n}\sum_c \sum_i \sum_j \epsilon_{A, i, c} \epsilon_{A, j, c} + \frac{1}{n} \sum_c \sum_i \sum_j \epsilon_{B, i, c} \epsilon_{B, j, c}
\]

We can denote the three terms on the right-hand side as

\[
\Var_{\rm clustered}(\hat{mu}_{A-B}) = \omega_{\rm clustered}^2 + \sigma_{A, {\rm clustered}}^2 + \sigma_{B, {\rm clustered}}^2
\]

with

\[
\omega_{\rm clustered}^2 = \Var_{\rm clustered}(x_A)+\Var_{\rm clustered}(x_B)-2\Cov_{\rm clustered}(x_A, x_B)
\]
\[
\Var_{\rm clustered}(x_A)=\frac{1}{n}\sum_c \sum_i \sum_j (x_{A, i, c}-\bar{s}_A)(x_{A, j, c}-\bar{s}_A)
\]
\[
\Var_{\rm clustered}(x_B)=\frac{1}{n}\sum_c \sum_i \sum_j (x_{B, i, c}-\bar{s}_B)(x_{B, j, c}-\bar{s}_B)
\]
\[
\Cov_{\rm clustered}(x_A, x_B)=\frac{1}{n}\sum_c \sum_i \sum_j (x_{A, i, c}-\bar{s}_A)(x_{B, j, c}-\bar{s}_B)
\]
\[
\sigma_{A, {\rm clustered}}^2 = \frac{1}{n} \sum_c \sum_i \sum_j \epsilon_{A, i, c} \epsilon_{A, j, c}
\]
\[
\sigma_{B, {\rm clustered}}^2 = \frac{1}{n} \sum_c \sum_i \sum_j \epsilon_{B, i, c} \epsilon_{B, j, c}
\]

These clustered versions of $\omega^2$, $\sigma_A^2$, and $\sigma_B^2$ can be plugged into Equations \ref{eqn:sample_size} and \ref{eqn:mde} without further modification.

In practice, in both the clustered and non-clustered cases, the mean conditional variance and variance of conditional means will need to be estimated from previous data having $K\gg 1$. For the sake of completeness, we briefly walk through this estimation process.

Let $s_{M, i, c, k}$ represent the $k$th score on the $i$th question within the $c$th cluster on model $M$ and estimate $\hat{x}_{M, i, c}=\frac{1}{K}\sum_{k=1}^K s_{M, i, c, k}$. This estimate is sufficient to estimate $\hat{\omega}_{\rm clustered}^2$. The clustered mean conditional variance on model $M$ may then be estimated as

\[
\hat{\sigma}_{M, {\rm clustered}}^2=\frac{1}{n(K-1)}\sum_k \sum_c \sum_i \sum_j (s_{M, i, c, k}-\hat{x}_{M, i, c})(s_{M, j, c, k}-\hat{x}_{M, j, c})
\]

Note that we divide by $K-1$ instead of $K$ in order to obtain a consistent variance estimator with small $K$.

If a subsample of questions is being used for variance estimation, we recommend sampling at the cluster level (i.e. drawing clusters in their entirety) in order to capture the intra-cluster variance structure.

\end{document}